\documentclass[aps,pre,twocolumn,showpacs,showkeys,superscriptaddress,floatfix]{revtex4}
 \usepackage{graphicx}

 \newcommand{\be}{\begin{equation}}
 \newcommand{\ee}{\end{equation}}
 \newcommand{\bea}{\begin{eqnarray}}
 \newcommand{\eea}{\end{eqnarray}}

\begin{document}

 \title{Convex lattice polygons of fixed area with perimeter dependent
weights}
 \author{R. Rajesh}
 \affiliation{Department of Physics - Theoretical Physics, University of
Oxford, 1 Keble Road, Oxford OX1 3NP, UK}
 \author{Deepak Dhar}
 \affiliation{Department of Theoretical Physics, Tata Institute of
Fundamental Research, Homi Bhabha Road, Mumbai 400005, India}
 \date{\today}

 \begin{abstract}
  We study fully convex polygons with a given area, and variable perimeter
length on square and hexagonal lattices. We attach a weight $t^m$ to a
convex polygon of perimeter $m$ and show that the sum of weights of all
polygons with a fixed area $s$ varies as $s^{-\theta_{conv}} e^{K
\sqrt{s}}$ for large $s$ and $t$ less than a critical threshold $t_c$,
where $K$ is a $t$-dependent constant, and $\theta_{conv}$ is a
critical exponent which does not change with $t$. We find
$\theta_{conv}$ is $ 1/4$ for the square lattice, but $-1/4$ for
the hexagonal lattice. The reason for this unexpected non-universality of
$\theta_{conv}$ is traced to existence of sharp corners in the
asymptotic shape of these polygons. 
 \end{abstract}
 \maketitle

\section{\label{sec1} Introduction}

The study of polygons is an important problem in lattice statistics
\cite{rensburg}.  It has been studied in the context of self-avoiding
walks, and as a model of the shape transition in vesicles
\cite{fisherguttmann1,fisherguttmann2}. The problem is also related to the
statistics of rare large clusters in percolation theory (see below). There
has been considerable progress in counting exactly various sub-classes of
polygons weighted by area and perimeter (see \cite{melou1,guttmann} and
references within). Recently, the exact critical scaling function of these
polygons has also been found \cite{RGJ,cardy,richard,RJG}. 

Convex polygons are an important sub-class of polygons. They are defined
as follows. The area enclosed by a polygon on a lattice is a simply
connected set of elementary plaquettes or cells of the lattice.  A polygon
on a square lattice is said to be column-convex if all the plaquettes in any
column are connected through plaquettes in the same column. The polygon is
convex if it is column-convex in both the horizontal and vertical directions
(see Fig.~\ref{fig1}). A polygon on a hexagonal lattice is said to be
convex if it is column-convex in all its three lattice directions (see
Fig.~\ref{fig2}).

Let $C_{m,s}$ be the number of convex polygons with perimeter $m$ and area
$s$.  We define the generating function
 \be
 C_s(t)= \sum_m C_{m,s} t^m. 
 \label{eq:1}
 \ee
 For any finite $s$, this is a finite polynomial, and hence convergent. 
For large $s$, there exists a $t_c<1$ such that for all $0<t<t_c$, the
leading contribution to the sum in Eq.~(\ref{eq:1}) comes from polygons
whose perimeter is of order $\sqrt{s}$. For the square lattice $t_c=1/2$.
It is straightforward to prove upper and lower bounds on this sum which
vary as an exponential of $\sqrt{s}$. It is expected that the leading
correction to the exponential behavior is a power law,
 \be
 C_s(t) \sim s^{-\theta_{conv}} e^{ K(t) \sqrt{s}},~s\rightarrow
\infty,~t < t_c,
 \label{eq:2}
 \ee
 where $K(t)$ is a $t$-dependent function, and $\theta_{conv}$ is a
critical exponent. When $t$ tends to zero, $K(t)$ tends to $C \ln(t)$,
where $C=4$ for the square lattice. The power-law 
exponent $\theta$ corresponding to
other sub-classes of polygons will be denoted by a suitable subscript.

In this paper, we calculate $\theta_{conv}$ for convex polygons on
the square and hexagonal lattices by summing over all polygons with a
fixed area and weighted by perimeter, and show that $\theta_{conv}$
for the square lattice is $1/4$, but for the hexagonal lattice it is
$-1/4$. We explain this difference by showing that the asymptotic shape of
large convex polygons on square and hexagonal lattices consist of $4$ and
$6$ cusps respectively. For a polygon whose macroscopic shape shows $n$
cusps, we argue that the value of $\theta$ is $(5-n)/4$. 

In the percolation problem (see \cite{stauffer1,stauffer2} for an
introduction)  above the percolation threshold, the probability
$\mbox{Prob}_p(s)$ of finite clusters of size $s$ in $d$-dimensions is
expected to vary as \cite{kunz,lubensky}
 \be \mbox{Prob}_p(s) \sim s^{-\theta_{perc}} \exp \left(- B(p)
{s}^{\frac{d-1}{d}} \right), \quad s\rightarrow \infty.
 \label{eq:3}
 \ee
 Here the exponent $\theta_{perc}$ is expected to be universal,
same for all $p$ above the critical percolation threshold.  For these rare
large clusters in two dimensions, the linear size of a cluster of $s$
sites varies as $\sqrt{s}$. It has a few holes, and the external boundary
of the cluster has overhangs. These are normally expected to be
irrelevant. On ignoring holes, we can model percolation clusters by
hole-less clusters, and $\mbox{Prob}_p(s)$ would have the same qualitative
behavior. In particular, $\theta_{perc} = \theta_{poly}$,
where $\theta_{poly}$ is the power-law exponent in Eq.~(\ref{eq:1})
corresponding to all lattice polygons with a fixed area. 

The macroscopic shape of rare large clusters for $p>p_c$ is convex.
Local fluctuations of the surface at a non-zero angle to
the $x$-axis can be well approximated by the fluctuations of a staircase
path. As most of the surface of the cluster has a non-zero finite slope, 
one may expect that dominant contribution to $\mbox{Prob}_p(s)$ comes
from convex polygons. This would imply that 
$\theta_{perc} = \theta_{poly}=\theta_{conv}$.
Our results show that both these equalities 
cannot be correct simultaneously. Presumably, the second equality is wrong
as $\theta_{conv}$ turns out to be lattice dependent.  For the
percolation problem, the presumably exact value of $\theta_{perc}$
has been calculated in all dimensions using techniques of continuum field
theory, within the droplet model which ignores the holes and overhangs in
the clusters \cite{lubensky}. In two dimensions $\theta_{perc}
=5/4$.

We now briefly review known results for convex polygons.  For convex
polygons on a square lattice, the exact two-variable generating function
$C(t,z)$ defined as
 \be
 C(t,z) = \sum_s C_s(t) z^s,
 \label{eq:5}
 \ee
 was calculated by Lin \cite{lin} and Bousquet-M\'{e}lou
\cite{melou2,melou3}. It was shown that
 \be
 C(t,z) = G + 2 \sum_{m=2}^{\infty} g_m \sum_{n=1}^{m-1} t^{-2 n}
\sum_{p=0}^{\infty} f_{n+p} +\sum_{m=3}^{\infty} g_m S_m,
 \label{eq:convex1}
 \ee
 where
 \begin{widetext}
 \bea
 g_m(t,z) &=&t^{2 m} \sum_{n=1}^{\infty} (t^2 z)^n \prod_{k=1}^n
(1-z^k)^{-2} \left[ u_{m-1,n} - (2+ z^n) u_{m-2,n} +(1+ 2 z^n) u_{m-3,n} -
z^n u_{m-4,n} \right], \nonumber\\
 u_{k,n}(t,z)&=& \sum_{r=0}^{k} \prod_{m=1}^{n+r} (1-z^k) 
\prod_{m=1}^{n+k-r} (1-z^k) \prod_{m=1}^{r} (1-z^k)^{-1} \prod_{m=1}^{k-r}
(1-z^k)^{-1},\quad k\geq 0, \nonumber\\
 G(t,z)&=& \sum_{m=1}^{\infty} g_m(t,z), \nonumber\\
 S_m(t,z) &=& \sum_{n=1}^{m-2} g_n t^{-2 n} (m-n-1), \label{eq:convex2}\\
 f_m(t,z) &=& h_m + \sum_{n=2}^{m} S_{n+1} \left[ \frac{t^2 z}{h'_1-h_1}
\left( h_m (h'_n-h_m) \right) +\delta_{m,n} t^{2 n+2} z^n \right],
\nonumber \\
 h_n(t,z) &=&t^{2 n+2} z^n \left( 1+ \sum_{m=1}^{\infty}\frac{(-t^2)^m
z^{m(m+1+2 n)/2}}{\prod_{r=1}^m (1-z^r) (1-t^2 z^r)} \right)  \left( 1+
\sum_{m=1}^{\infty}\frac{(-t^2)^m z^{m(m+1)/2}}{\prod_{r=1}^m (1-z^r) 
(1-t^2 z^r)} \right)^{-1}, \nonumber \\
 h'_n(t,z) &=&t^2 z^n \left( 1+ \sum_{m=1}^{\infty}\frac{(-t^2)^m
z^{m(m+1+2 n)/2}}{\prod_{r=1}^m (1-z^r) (t^2- z^r)} \right)  \left( 1+
\sum_{m=1}^{\infty}\frac{(-t^2)^m z^{m(m+1)/2}}{\prod_{r=1}^m (1-z^r)
(t^2- z^r)} \right)^{-1}. \nonumber
 \eea
 \end{widetext}
 It is not easy to extract the asymptotic behavior of $C_s(t)$ for large
$s$ and fixed small value of $t$ from the complicated expressions
Eqs.~(\ref{eq:convex1}) and (\ref{eq:convex2}). 

The asymptotic behavior of the coefficient of $t^m$ in Eq.~(\ref{eq:5}),
when $z>1$, was determined in Ref.~\cite{prellberg}. In this case, the
dominant contribution comes from the largest $s$ possible, which is
$z^{m^2/16}$ \cite{fisherguttmann2}.  To be more specific, it was proved
\cite{prellberg} that for fixed $z>1$
 \be
 \sum_s C_{m,s} z^s = A(z) z^{m^2/16} (1+(\rho^m)), ~m \rightarrow \infty,
 \label{eq:6}
 \ee
 for some $\rho<1$. The function $A(z)$ was shown to behave as
 \be
 A(z) \sim \frac{1}{4}\left(\frac{\epsilon}{2 \pi}\right)^{3/2} e^{2
\pi^2/(3 \epsilon)} ~\mbox{as}~\epsilon = \ln(z) \rightarrow 0^+.
 \label{eq:7}
 \ee
 These results are valid when $\epsilon m \gg 1$.  We are interested in
the case when $\epsilon \rightarrow 0^-$ with $m \sim \sqrt{s} \sim
1/\epsilon$. It is not clear how to extend the results Eqs.~(\ref{eq:6}) 
and (\ref{eq:7}) in this regime.  However, if we {\it assume} that the
results remain valid qualitatively in this regime also, and the limits of
$m$ large and $\epsilon$ small can be taken in reverse order, we can
estimate $C_{s,t}$ by
 \be
 C_{s}(t)= \sum_m t^m \frac{1}{2 \pi i}\oint \frac{d z}{z^{s+1}} \sum_s
C_{m,s} z^s. 
 \label{eq:contour}
 \ee
 Equation~(\ref{eq:contour}) can be evaluated by the method of steepest
descent giving $C_{s}(t) \sim s^{-5/4} e^{\sqrt{s} K(t)}$. This conclusion
seems to be correct for all polygons, but as we shall show later in the
paper, it is incorrect for convex polygons. This implies that in the
regime of interest, the asymptotic behavior is indeed different and not
given by Eqs.~(\ref{eq:6}) and (\ref{eq:7}).

The rest of the paper is organized is as follows. In Sec.~\ref{sec2}, the
exponent $\theta_{conv}$ is calculated for the square and hexagonal
lattice. In Sec.~\ref{sec3}, the macroscopic shape of convex polygons is
determined. In Sec.~\ref{sec4}, the results are extended to sub-classes of
convex polygons. In Sec.~\ref{sec5}, the macroscopic shape of
column-convex polygons is determined.  Finally, we end with a summary and
conclusions in Sec.~\ref{sec6}.

\section{\label{sec2}Calculation of the exponent $\theta_{conv}$}

\begin{figure}
 \includegraphics[width=8.0cm]{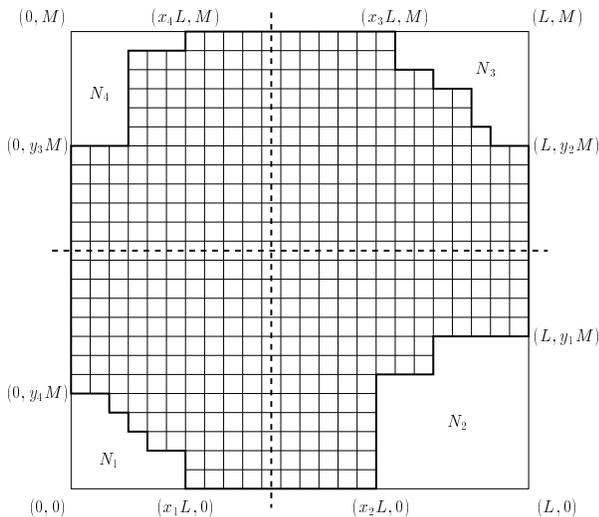}
 \caption{\label{fig1}A typical convex polygon on a square lattice and its
bounding box is shown. All vertical and horizontal straight lines (dotted
in the figure) intersect the polygon either $0$ or $2$ times.  The convex
polygon can be thought of as a rectangle from whose corners some squares
have been removed by staircase like paths.}
 \end{figure}

Consider convex polygons on a square lattice.  A convex polygon of a given
perimeter can be visualized as a bounding rectangle of the same perimeter 
from each
of whose four corners squares have been removed by staircase like paths
(see Fig.~\ref{fig1}). Such paths are also known as Ferrers diagrams in
the combinatorics literature.  These staircase paths have the constraint
that they cannot intersect each other.  All convex polygons may then be
generated by considering all possible rectangles. 

Let $R(z,A,B)$ be a generating function such that the coefficient of $z^s$
enumerates the number of staircase paths from $(0,A)$ to $(B, 0)$
enclosing an area $s$. We then obtain
 \bea
 \lefteqn{ \sum_s C_s(t) z^s \sim \sum_{x_i, y_i, L, M} t^{2 (L+M)} z^{L
M}} \nonumber \\
 && \!\!\!\!\!\!\!\! \times 
R(z^{-1}\!, x_1 L, y_4 M)  ~ R(z^{-1}\!, (1-x_3) L, (1-y_2) M)
\nonumber \\
 && \!\!\!\!\!\!\!\! \times
 R(z^{-1}\!, (1-x_2) L, y_1 M)  ~ R (z^{-1}\!, x_4 L, (1-y_3) M),
 \label{eq:8}
 \eea
 where we refer to Fig.~\ref{fig1} for the notation.  In writing down
Eq.~(\ref{eq:8})  we have ignored the case when the staircases at two
opposite corners may intersect. This will only make an exponentially small
correction and will not modify the exponent $\theta_{conv}$. From
the theory of partitions \cite{andrews}, it is known that
 \be
 R(z, A, B)= z^{A+B-1} \frac{(z)_{A+B-2}}{(z)_{A-1} (z)_{B-1}},
 \label{eq:9}
 \ee
 where
 \be
 (z)_A= \prod_{k=1}^{A}(1-z^k). 
 \label{eq:10}
 \ee

The asymptotic behavior of the coefficient of $z^s$ in $R(z,A,B)$ for
large $s$ can be calculated by the method of steepest descent. To evaluate
$(z)_A$, we take logarithms on both sides of Eq.~(\ref{eq:10}) and convert
the resultant sum into an integral by using the Euler-Maclaurin sum
formula \cite{bender}. This gives
 \be
 (z)_{A} \sim \frac{1}{\sqrt{\epsilon}} \exp \left(\frac{1}{\epsilon}
\int_{e^{-\epsilon A}}^1 \frac{\ln(1-x)}{x} \right), ~~ \epsilon =-\ln(z)
\rightarrow 0.
 \ee
 We would be interested in the limit when $A$ and $B$ vary as $\sqrt{s}$.
Let $A=a\sqrt{s}$ and $B=b \sqrt{s}$. Then, the coefficient of $z^s$ in
$R(z,A,B)$ is given by
 \be
 \frac{1}{2 \pi i} \oint \frac{R(z,A,B)}{z^{s+1}} \sim
\frac{1}{s^{3/4}} \int d \alpha e^{\sqrt{s} g(\alpha,a,b)},
 \label{eq:coeff_R}
 \ee
 where we made the substitution $z=e^{-\alpha/\sqrt{s}}$, and the
function  $g(\alpha,a,b)$ is given by
 \bea
\lefteqn{ g(\alpha,a,b) = 
\alpha +\frac{1}{\alpha} \bigg( \int_{e^{-\alpha(a+b)}}^{1}
du \frac{\ln(1-u)}{u}} \nonumber \\
 && \mbox{}- \int_{e^{-\alpha a}}^{1} du \frac{\ln(1-u)}{u} 
- \int_{e^{-\alpha b}}^{1} du \frac{\ln(1-u)}{u} \bigg).
 \label{eq:g}
 \eea
 The function $g(\alpha,a,b)$ has has a minimum at some
$\alpha=\alpha_0$ where $\alpha_0$ is a function of $a$ and $b$. On
doing the integral in Eq.~(\ref{eq:coeff_R}) by the method of steepest
descent, the power law factor gets modified by a
factor $s^{-1/4}$. Thus, we obtain
 \be
 \frac{1}{2 \pi i} \oint \frac{R(z,A,B)}{z^{s+1}} \sim \frac{1}{s}
\exp\left[ \sqrt{s} f(A/\sqrt{s}, B/\sqrt{s}) \right],
 \ee
 where
 \be
 f(a,b) = g(\alpha_0,a,b),
 \ee
 with $g(\alpha,a,b)$ as in Eq.~(\ref{eq:g}) and $\alpha_0$ being that
value of $\alpha$ which minimizes $g(\alpha,a,b)$.
The function $f(a,a)$ increases monotonically from $0$ to $\pi
\sqrt{2/3}$ when $a$ increases from $1$ to $\infty$. The value at
infinity, $f(\infty,\infty)$, corresponds to the result for unrestricted
partitions \cite{hardyramanujan}. Clearly, the function $f(a,b)$ is a
monotonically increasing function in both its variables.

From now onwards, we will consider the case when all the distances in
Fig.~\ref{fig1} scales as $\sqrt{s}$, i.e., $L=l \sqrt{s}$, and $M=m
\sqrt{s}$. Also, each of the $N_i$'s varies linearly with $s$, i.e.,
$N_i=n_i s$. Equation~(\ref{eq:8}) then reduces to
 \begin{widetext}
 \bea
 \lefteqn{ C_s(t) \sim \int \prod_{i=1}^4 (dx_i dy_i dn_i) dl dm
\left(\sqrt{s} \right)^{10} s^4 \delta \left[s \left(1+\sum n_i-l
m)\right)\right] t^{2
 \sqrt{s}(l+m)}} \nonumber \\
 && \times \frac{1}{s^4} \exp\left(\sqrt{s}\left[ \sqrt{n_1}
f\left(\frac{x_1 l}{\sqrt{n_1}}, \frac{y_4 m}{\sqrt{n_1}}\right)  +
\sqrt{n_2} f\left(\frac{(1-x_2) l}{\sqrt{n_2}}, \frac{y_1
m}{\sqrt{n_2}}\right)  \right] \right) \nonumber \\ && \times
\exp\left(\sqrt{s}\left[ \sqrt{n_3} f\left(\frac{(1-x_3) l}{\sqrt{n_3}},
\frac{(1-y_2) m}{\sqrt{n_3}}\right)  + \sqrt{n_4} f\left(\frac{x_4
l}{\sqrt{n_4}}, \frac{(1-y_3) m}{\sqrt{n_4}}\right)\right] \right),
 \label{eq:11}
 \eea
 \end{widetext}
 where the $(\sqrt{s}))^{10}$ factor is due to the scaling of the
distances, the $s^4$ factor is due to scaling of the $N_i$'s and $s^{-4}$
factor is due to the power law term in the asymptotic formula for
partitions. Thus, there is an overall power law factor $s^{5}$.

In the limit of large $s$, the integrals can be performed by the saddle
point method. We first note that the shape that has maximum contribution
to the integral will have the symmetry of the square lattice, i.e., the
bounding box will be a square of side $x_0 \sqrt{s}$ and each of the
$N_i$'s will be equal to $\beta_0 s$. Consider the integration over the
variables $x_2, y_2, x_3, y_3$ about this shape.  Due to the monotonic
behavior of the scaling function $f(x,y)$, the integrand is maximum with
respect to these four variables at the end points of their limits of
integration, namely $x_1, y_1, x_3, y_3$ respectively. On doing the saddle
point integration, the contribution to $\theta_{conv}$ in the power
law prefactor from each integration is $s^{-1/2}$. Thus, we are left with
a power law term $s^3$.  With respect to the remaining coordinate
variables $x_1, y_1, x_3, y_3, l, m$, the integrand takes on it maximum
value at a point in the interior of the region of integration, and each
such integration contributes a factor $s^{-1/4}$ to the power law
prefactor. Thus, after integrating over all the coordinates, a power law
factor of $s^{3/2}$ remains.

Now, only the integrals over the $n_i$'s remain to be done. Out of the
four integrals, one of them integrates away the delta function
contributing a factor $s^{-1}$, while each of the others contributes a
factor $s^{-1/4}$ to the power law prefactor.  Collecting together these
terms, we obtain
 \be
 C_{s}(t) \sim \frac{1}{s^{1/4}} e^{4 \sqrt{s} \left( \sqrt{\beta_0}
f\left( \frac{x_0}{2\sqrt{\beta_0}}, \frac{x_0}{2\sqrt{\beta_0}} \right) 
+x_0 \ln(t)\right)}.
 \label{eq:12}
 \ee
 We compute the term in the exponential in Eq.~(\ref{eq:12}) in
Sec.~\ref{sec3} (see Eq.~(\ref{eq:21})). Equation~(\ref{eq:12}) implies
that for convex polygons on a square lattice
 \be
 \theta_{conv}^{sq}= \frac{1}{4}. 
 \label{eq:13}
 \ee
 \begin{figure}
 \includegraphics[width=5.0cm]{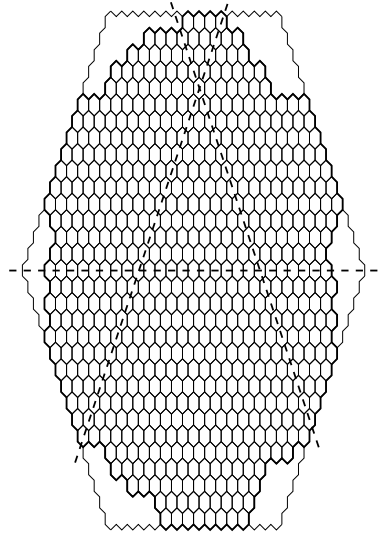}
 \caption{\label{fig2}A typical convex polygon on a hexagonal lattice is
shown. Any straight line in the three lattice directions (shown as dotted
lines)  intersect the polygon at most twice. The convex polygon can be
thought of 6 blocks carved out by directed staircase like paths from a
bounding hexagon.}
 \end{figure}

The above calculation of $\theta_{conv}$ can be summarized as
follows. Consider a convex polygon constructed from a bounding box by $n$
staircase paths ($n=4$ for square lattice). The end points of the
staircase paths can slide along the bounding box, and each path
contributes three coordinates to be integrated over. Out of these $3n$
coordinates, two of them are fixed to prevent over counting of polygons
which are identical modulo translations. Thus, there are a total of $(3 n
-2)$ coordinates, each one of them varying as $\sqrt{s}$, to be integrated
over. Each staircase path also encloses an area, varying as $s$, that has
to be integrated over.  Finally, there is a contribution $s^{-1}$ from
each such area, corresponding to the enumeration of staircase paths with
fixed ends and fixed area. Thus, the integrand has an overall power law
factor $s^{(3 n -2)/2}$ to start with. On doing the integrations, the
first $n$ coordinate integrals contribute a factor $s^{-1/2}$ each as the
maximum occurs at the end points of the integration limits, while the
remaining $(2 n - 2)$ coordinates contribute $s^{-1/4}$ each. Thus, after
the integration over the coordinates, the power law factor is
$s^{(n-1)/2}$. The integrations over the areas have the following
contributions. One of them integrates over the delta function,
contributing $s^{-1}$, while each of the other contribute a factor
$s^{-1/4}$. Taking these corrections into account, we obtain that
$\theta_{conv}$ for a $n$-sided convex polygon is
 \be
 \theta_{conv}^{n\mbox{-}sided} = \frac{5-n}{4}. 
 \label{eq:14}
 \ee
 We recover the square lattice result (Eq.~(\ref{eq:13})) when $n=4$ in
Eq.~(\ref{eq:14})

Consider now convex polygons on a hexagonal lattice (see Fig.~\ref{fig2}). 
It is quite straightforward to carry out a similar analysis as was done
for the square lattice. Equivalently, putting $n=6$ in the expression for
$\theta_{conv}$ for $n$-sided convex polygons, we obtain
 \be
 \theta_{conv}^{hex}=-\frac{1}{4}. 
 \label{eq:15}
 \ee
 Equations~(\ref{eq:13}) and (\ref{eq:15}) imply that
$\theta_{conv}$ is not universal for convex polygons and takes on
different values on different lattices. 

\section{\label{sec3}Macroscopic shape of convex polygons}

The fact that $\theta_{conv}$ for the square and hexagonal lattices
comes out different is somewhat unexpected. To understand the reason for
this difference, and why this differs from the value 
$\theta_{perc} =5/4$ for
percolation clusters, we need to look at the macroscopic shape of convex
polygons.  This can be done exactly using the Wulff construction
\cite{wulff}. Consider the case on the square lattice. The equilibrium
curve is the one that extremises the free energy functional
 \be
 {\mathcal{L}}[y(x)]=\int_0^{X} dx \sigma (y^\prime)  \sqrt{1+y^{\prime
2}} -\frac{2 \lambda}{\sqrt{s}} \int_0^X y dx,
 \label{eq:16}
 \ee
 where $y^\prime=dy/dx$, $\sigma(y^\prime)$ is the orientation dependent
surface tension and $\lambda$ is a Lagrange multiplier. The equilibrium
curve $y_0(x)$ satisfies the Euler-Lagrange equation
 \be
 -\frac{d}{dx} \left( \frac{d}{d
y^\prime}\left(\sigma(y^\prime)\sqrt{1+y^{\prime 2}}\right)\right)  -
\frac{2 \lambda}{\sqrt{s}}=0.
 \label{eq:17}
 \ee
 The equilibrium macroscopic shape is then obtained by minimizing
${\mathcal{L}}[y_0(x)]$ with respect to the endpoint $X$.

 For convex polygons, it is easy to determine the slope dependent surface
tension exactly. It has two contributions: one coming from the energy of
the interface, and one from the entropy. For an interface having $X$
horizontal and $Y$ vertical steps, the energy per unit length is
proportional to $|X| + |Y |$, and the number of configurations is $
(|X|+|Y|)!/|X|!|Y|!$. This gives
 \bea
 \sigma(y') \sqrt{1+y^{\prime 2}} &=& -(1+|y'|) \ln(1+|y'|)  +|y'|
\ln(|y'|) \nonumber\\ &&\mbox{}- (1+|y'|) \ln (t).
 \label{eq:18}
 \eea
 Following the above procedure, we obtain that the macroscopic shape of
the staircase satisfies the equation
 \be
  e^{- 2 \lambda |y|/\sqrt{s}}+ e^{-2 \lambda |x|/\sqrt{s}}= t^{-1},
 \label{eq:19}
 \ee
 where the Lagrange multiplier $\lambda$ is the negative root of
 \be
 \lambda^2 = \ln(t) \ln\left(\frac{t}{1-t}\right) +\int_{t}^{1-t} du
\frac{\ln(1-u)}{u}. 
 \label{eq:20}
 \ee
 At $t$ tends to zero, the shape tends to the square $\max(|x|,|y|) =
\sqrt{s}/2$. When $t$ tends to $1/2$, then $\lambda$ tends to zero and the
shape tends to $|x| + |y| = \sqrt{s/2}$. When $t=1$, The shape
Eq.~(\ref{eq:19}) reduces to that for unrestricted partitions
\cite{temperley1,vershik}. 

\begin{figure}
 \includegraphics[width=8.0cm]{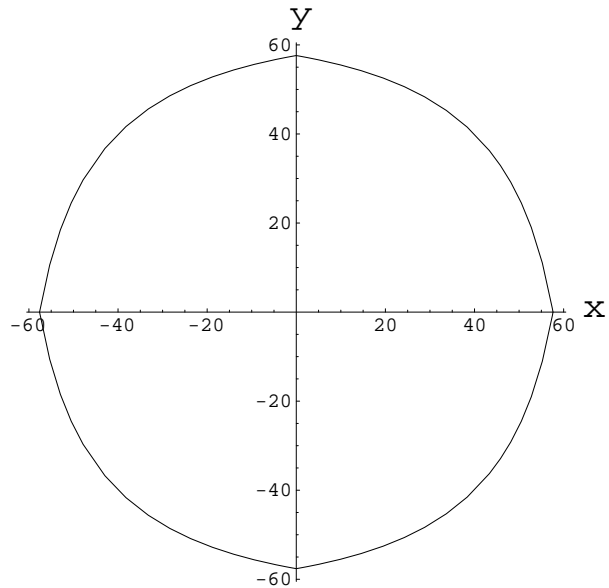}
 \caption{\label{fig4}
 The equilibrium shape of a convex polygon on a square lattice enclosing
an area $10000$ when the perimeter weight $t=0.15$ is shown.}
 \end{figure}

The term in the exponential of Eq.~(\ref{eq:12}) can be calculated by
substituting Eqs.~(\ref{eq:18}), (\ref{eq:19}) and (\ref{eq:20}) into
Eq.~(\ref{eq:17}). Doing so, we obtain
 \be
 C_s(t) \sim \frac{1}{s^{1/4}} e^{\lambda \sqrt{s}},
 \label{eq:21}
 \ee
 where $\lambda$ is a function of $t$ determined by Eq.~(\ref{eq:20}). 
The equilibrium shape of a convex polygon enclosing an area $10000$ when
$t=0.15$ is shown in Fig.~\ref{fig4}. The four staircase paths intersect
each other at a finite angle. The reason why we see cusps in the
macroscopic shape is the term proportional to $|y'| \ln(|y'|)$ in the
expression for the direction dependent surface tension $\sigma(y')$.  This
singular term makes $\sigma(y')$ a local maximum at $y'=0$, which leads to
a cusp. The macroscopic shape has four cusps due to the four-fold symmetry
of the square lattice.

A similar analysis can be done for convex polygons on a hexagonal lattice. 
The surface energy $\sigma(y')$ has qualitatively the same behavior as for
the square lattice. The six fold symmetry of the hexagonal lattice results
in $6$ cusps for the hexagonal convex polygons. 

For ordinary percolation, the continuum theory calculation \cite{lubensky}
gives $\theta_{perc} = 5/4$. 
On the other hand, $\theta_{conv}$ for a
$n$-sided convex polygon takes on the value $5/4-n/4$. In addition, the
macroscopic shape of a $n$-sided convex polygon has $n$ cusps. These cusps
are not expected to appear in the macroscopic shape of percolation
clusters.  One would presume that on going beyond the convex polygons
approximation, these cusps would disappear, each contributing a certain
factor to the power law. Thus, putting $n=0$ in Eq.~(\ref{eq:14}), we
recover the result for percolation.

We can similarly determine the value of $\theta_{dir~perc}$ 
for two dimensional directed percolation (see
\cite{hinrichsen} for definition and an introduction). Consider
directed percolation above the percolation threshold. Let the infinite
cluster have a finite opening angle $\pi/2- 2 \gamma$, where $\gamma$ is a
function of $p$. Then, the surface tension for surfaces which have slopes
$\tan(\gamma)$ and $\tan(3 \pi/2 - \gamma)$ is zero. Due to these local
minima, and hence a maximum at zero slope, the macroscopic shape of finite
directed percolation has a cusp at the origin with an opening angle
$\pi/2- 2 \gamma$. Thus, $\theta_{dir~perc}$ for directed
percolation is obtained by substituting $n=1$ in Eq.~(\ref{eq:14}),
yielding
 \be
 \theta_{dir~perc} = 1 \quad \mbox{in $2$-dimensions}. 
 \label{eq:23}
 \ee

\section{\label{sec4}Sub-classes of convex polygons}

\begin{figure}
 \includegraphics[width=8.5cm]{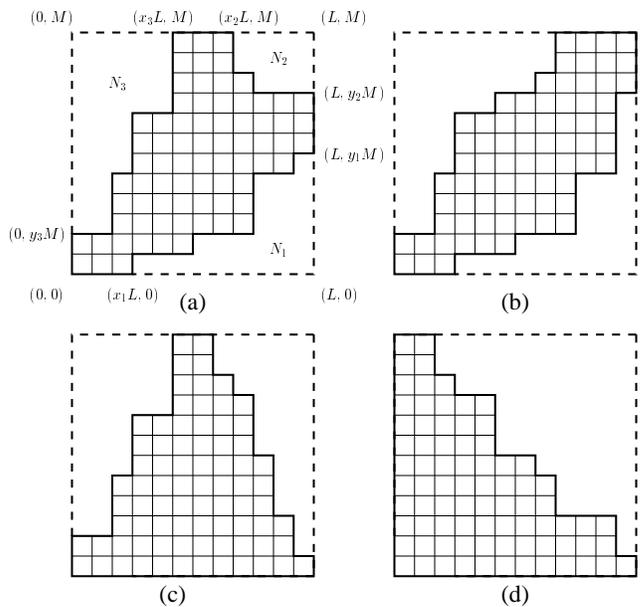}
 \caption{\label{fig5} Examples of (a) a directed convex polygon, (b) a
staircase polygon, (c) a pyramidal polygon and (d) Ferrers diagram on a
square lattice.}
 \end{figure}

In this section, we extend the results to sub-classes of convex polygons.
A directed convex polygon on a square lattice is a convex polygon for
which the lower left corner of the bounding rectangle is also a vertex of
the polygon (see Fig.~\ref{fig5}(a)). As for convex polygons, the area and
perimeter weighted generating function for directed convex polygons is
known \cite{melou2,melou3}. We now determine the exponent $\theta$ in
exactly the same way as was done for convex polygons. 

Consider a directed convex polygon.  The contribution to the power law
prefactor from the various steps in the power counting is as follows.  (1)
The integrand initially has a power law factor $(\sqrt{s})^8$.  (2)
Integration over $x_1, y_1, x_3, y_3$ contributes $s^{-1/2}$ each to the
power law factor. (3) Integration over $x_2, y_2, l, m$ contributes
$s^{-1/4}$ each to the power law factor. (4) Integration over
$N_1,N_2,N_3$ contributes $(s^{-1/4})^2 s^{-1}$ to the power law factor. 
Collecting together the various terms, we obtain
 \be
 \theta_{dir~conv} = \frac{1}{2},
 \label{eq:22}
 \ee
 where $\theta_{dir~conv}$ is the $\theta$ corresponding to
directed convex polygons. The macroscopic shape of the directed convex
polygon has three cusps. Not surprisingly, substituting $n=3$ in
Eq.~(\ref{eq:14})  gives the result in Eq.~(\ref{eq:22}). 

The other sub-classes of convex polygons that we study are staircase
polygons, pyramidal polygons and Ferrers diagrams. Staircase polygons are
convex polygons for which both the lower left and upper right corners of
the bounding rectangle are vertices of the polygon.  Pyramidal polygons
are convex polygons for which both the lower left and lower right corners
of the bounding rectangle are vertices of the polygon. Ferrers diagrams
are convex polygons for which the lower left, lower right and upper left
corners of the bounding rectangle are vertices of the polygons. Examples
of the polygons are shown in Fig.~\ref{fig5}(b), \ref{fig5}(c) and
\ref{fig5}(d) respectively. The area and perimeter generating function of
staircase polygons \cite{polya,brak1}, pyramidal polygons
\cite{temperley1} and Ferrers diagram \cite{andrews} are known. The
exponent $\theta$ can be calculated for each one of them as before. 
Proceeding on the same lines, we obtain
 \bea
 \theta_{stair} &=& \frac{3}{4},\\
 \theta_{pyramid} &=& \frac{1}{2},\\
 \theta_{Ferrer} &=& \frac{1}{2}. 
 \eea
 These correspond to $2$, $3$ and $3$ cusps respectively in the
macroscopic equilibrium shape of these polygons. 

\section{\label{sec5}Column convex polygons}

\begin{figure}
 \includegraphics[width=8.5cm]{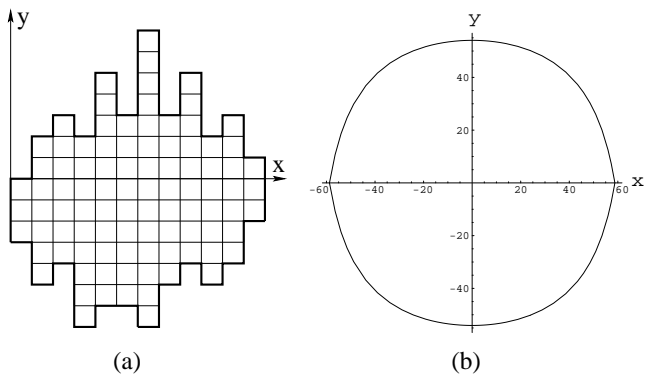}
 \caption{\label{fig6} (a) A column-convex polygon on a square lattice. 
Any line in the vertical direction intersects the polygon at either zero
or two points.  (b) The equilibrium macroscopic shape of a column-convex
polygon on a square lattice enclosing an area $10000$, when $t=0.15$.}
 \end{figure}

In this section, we determine the equilibrium shape of column-convex
polygons and show that it has two cusps. 
An example of a column-convex polygon is shown in
Fig.~\ref{fig6}(a). The area and perimeter weighted generating function
for column-convex polygons is known \cite{brak1}. However, as for convex
polygons, it is difficult to extract from it the asymptotic behavior of
fixed area polygons.

We first calculate the angle dependent surface tension $\sigma_r(\gamma)$,
where $y'=\tan(\gamma)$, for column-convex polygons.  This analysis is
similar to that done for directed polymers \cite{rajesh_poly}. Consider
all possible directed walks from $(0,0)$ to $(x,y)$. Then, the sum over
all weighted paths is
 \be
 e^{- x \sec(\gamma) \sigma_r(\gamma)} = \sum_{y_1,\ldots,y_x}
\delta\left(\sum_{i=1}^{x} y_i -y\right)  \prod_{i=1}^x t^{1+ | y_i |},
 \ee
 where $\delta$ is the usual Kronecker delta function. Taking Laplace
transform with respect to $y$, we obtain independent summations over
$y_i$. These are easily done giving
 \be
 \sigma_r(y') \sqrt{1+y'^2}= y'\ln(z_0)+\ln\frac{(1-t z_0) (1-t z_0^{-1})}
{t  (1-t^2)},
 \label{eq:row_free}
 \ee
 where
 \be
 z_0=\frac{(1+t^2)y' + \sqrt{(1-t^2)^2 y'^2 + 4 t^2}} {2 t(1+y')}. 
 \ee
 We see that $\sigma_r(y')$ is now a smooth function of $y'$ for $y'$ near
zero. For convex polygons, a surface with average orientation $y'=0$
cannot have any fluctuations, as the height fluctuations in the
$y$-direction become the disallowed overhangs in the $x$-direction. This
leads to the singularity near $y'=0$ in the expression for orientation
dependent surface tension for convex polygons. 

To construct the equilibrium shape of the polygon, we need to find the
$y(x)$ satisfying the Euler Lagrange equation (see Eq.~(\ref{eq:17})) with
$\sigma_r$ and a Lagrange multiplier $\lambda_r$. The curve $y(x)$
satisfies the boundary condition $y(-X/2)=0$ and $y(X/2)=0$. Solving, we find
that the shape of the polygon is given by
 \be
 e^{2 \lambda_r y/\sqrt{s}}= 4 c t 
\sinh\left(\frac{\ln(t)}{2} - \frac{\lambda_r x}{\sqrt{s}}\right) 
\sinh\left(\frac{\ln(t)}{2} + \frac{\lambda_r x}{\sqrt{s}}\right),
 \label{eq:row_shape}
 \ee
 where $c$ is a constant, $X=g(c) \sqrt{s}/\lambda_r$ and
 \be
 g(c) = \ln \left[ \frac{c(1+t^2) -1 + \sqrt{(1-c (1-t^2))^2 -4 c^2 t^2}}
{ 2 c t} \right].
 \ee 
The Lagrange multiplier $\lambda_r$ is fixed by the constraint that
$\int_{-X/2}^{X/2} y dx=s/2$. We obtain $\lambda_r$ as a function of $c$ 
to be
 \be
 \lambda_r^2 = \int_0^{g(c)} dz \ln\left[c (1-t e^{-z}) (1-t e^{z})\right].
 \label{eq:row_lambda}
 \ee
 The value of $c$ is chosen to be the one that minimizes the total
surface free energy. For the curve Eq.~(\ref{eq:row_shape}), the total
surface energy $F(c,t)$ is
 \be
 F(c,t)= 
2 \lambda_r \sqrt{s}
-\frac{ \sqrt{s}}{\lambda_r} g(c) \ln\left[c t (1-t^2) \right].
 \label{eq:row_energy}
 \ee
 Minimizing Eq.~(\ref{eq:row_energy}) with respect to $c$, we obtain
 \be
 c = \frac{1}{t (1-t^2)}. 
 \label{eq:row_c1}
 \ee

Equations~(\ref{eq:row_shape}), (\ref{eq:row_lambda}) and
(\ref{eq:row_c1}) describe the equilibrium shape of column-convex
polygons. In Fig.~\ref{fig6}(b), the shape when $t=0.15$ is shown. It has
two cusps. Thus, we would conclude from Eq.~(\ref{eq:14}) that
 \be
 \theta_{col~conv} = \frac{3}{4}. 
 \ee
 The height fluctuations of the column-convex polygons become overhangs 
when viewed after rotation by $\pi/2$. 
On introducing such overhangs, two of the four cusps
that were present in the shape of convex polygons vanished. Thus, one
would expect that if overhangs in the horizontal direction were also
allowed as in self avoiding polygons, then there would be no cusps, and
 \be
 \theta_{poly} = \frac{5}{4}. 
 \ee

Finally, we note that the macroscopic shape of column-convex polygons
becomes unstable when $\sigma_r(0)=0$. The smallest absolute value of $t$ at 
which this occurs is
 \be
 t_c = \sqrt{2} -1. 
 \ee
 This value matches with the previously obtained value for $t_c$
\cite{temperley2,brak2}. 

\section{\label{sec6}Summary and conclusion}

To summarize, we studied fixed area convex polygons weighted by their
perimeter on square and hexagonal lattices. The exponent
$\theta_{conv}$ as defined in Eq.~(\ref{eq:2}) was found to be
$1/4$ for the square lattice and $-1/4$ for the hexagonal lattice. This
discrepancy was traced to the presence of cusps in the macroscopic shape
of convex polygons. We argued that for a polygon whose macroscopic shape
has $n$ cusps has $\theta_n= (5-n)/4$. For polygons, one expects that the
macroscopic shape has no cusps.  Indeed, putting $n=0$ in
Eq.~(\ref{eq:14}), we recover the result $\theta_{perc}=5/4$
obtained for percolation clusters \cite{lubensky}. For directed
percolation, it is argued that there should be one cusp, and hence
$\theta_{dir~perc}=1$ in two dimensions.

\section*{Acknowledgments}

R. Rajesh would like to acknowledge EPSRC, U.K. for financial support.

\end{document}